# High-pressure formation and stabilization of iridium hydrides

Patryk Zaleski-Ejgierd [a]

[a] Institute of Physical Chemistry, Polish Academy of Sciences,
ul. M. Kasprzaka 44/52, 02-144, Warszawa, Polska

A class of iridium hydrides, $IrH_x$ ($x$=0.5, 1-4, 6), is calculated at *ab-initio* level using the DFT. A number of hydrides are predicted to stabilize in the excess hydrogen environment and upon compression. The threshold pressure for thermodynamical stabilization is calculated to ~5 and ~14 GPa at which, respectively, a trihydride ($P6_3mc$) and a dihydride ($Ibam$) stabilize with respect to decomposition into pure element. The dihydride ($Ibam$) is calculated to stabilize dynamically at P~14 GPa and the trihydride at P > 25 GPa. In both the di- and trihydride iridium atoms form short directional contacts with hydrogens, typical for covalent bonds. The trihydride exhibit molecular character and its relative stability is calculated to be higher than that of the dihydride, at least up to approx. 75 GPa.

## I. INTRODUCTION

The very first transition metal hydride, that of palladium, was synthesized already in 1866 [1] but it took over a century to synthesize another. The reason for it is that although the chemical potential of hydrogen dissolved in a transition metal can be actually very high, the solubilities of hydrogen, corresponding to equivalent high external pressures of the hydrogen gas, are generally very low, resulting in its typical inertness at ambient pressure. To increase the reactiveness one thus needs to substantially increase the hydrogen chemical potential.

The most direct way is to compress the hydrogen to very high pressures. Indeed, after nearly 100 years since the first report on palladium hydride formation, Baranowski *et. al.* were able to synthesize nickel hydride by simply increasing the pressure of hydrogen to ~15 kbar (1.5 GPa) [2-3]. Further extensions of the technique allowed for subsequent synthesis of many other transition metal hydrides.

Most of the d-block metals were found to form primarily non-stoichiometric hydrides in which hydrogen atoms are incorporated into the metal host lattice [4]. Of the transition metal hydrides only a few exhibit stoichiometric, or near-stoichiometric compositions: 2:1 ($Mn_2H$), 1:1 (WH, CrH, HfH) [5], or 1:2 ($ZrH_2$, $ZnH_2$). Two notable exception are scandium and yttrium which react with hydrogen at rare 1:3 ratio to form $ScH_3$ and $YH_3$; note that while the two metals formally belong to *d*-block they actually qualify more as rare-earth metals rather than true transition metals.

With few exception the remaining majority of transition metals form interstitial hydrides in a wide range of non-stoichiometric compositions. In fact, many of those "hydrides" should actually be considered solid-state hydrogen solutions rather than actual hydrides. Metals such as Cr, Mo, Mn, Tc, Re, Fe and Co were found to form hydrides based primarily on the *hcp* arrangement of the metal lattice while i.e. Pd is known to form hydrides based mainly on the *fcc* symmetry [6]. Other notable examples include Ti in which the host metal lattice adopts *bcc* symmetry [7-8]. In some cases, the available lattice vacancies may be occupied either randomly ($ReH_{0.23}$, $FeD_{0.42}$) [9-10] or in a superstructure order resulting in complex crystal structures, as in the case of $TcH_x$ hydrides [11-12].

Within Group-9 only cobalt and rhodium are known to form hydrides.

Cobalt forms non-stoichiometric hydrides, $CoH_x$, already at pressures as low as 4 GPa. For $x \leq 0.26$ the hydrogen atoms were found to be randomly distributed over the available octahedral sites of the *hcp* host metal lattice. For $x \geq 0.34$ the resulting cobalt hydride forms complex layered structures. With further pressure increase hydrogen concentration increases monotonically until it reaches $x \sim 0.6$ at 7 GPa. At even higher pressures a near-stoichiometric *fcc*-based hydride forms with $x \sim 1.0$ [13]. No higher hydrides in solid-state are known for cobalt. Interestingly, cobalt hydride can be retained in a metastable state at atmospheric pressure and low temperatures.

Elemental rhodium adopts the *fcc* crystal structure and as such it has one octahedral and two tetrahedral sites available per each Rh atom. A stable monohydride, RhH, is known in which one octahedral site of the *fcc* host lattice is occupied [14-15]. Recently Li *et. al.* reported high-pressure synthesis of a dihydride, $RhH_2$, in which two *tetrahedral* sites are occupied instead. This dihydride forms spontaneously from RhH in the excess hydrogen environment already at a relatively low pressure of ~8-10 GPa [16]

Iridium remains the most reluctant hydrogen acceptor among Group-9 and, in fact, among all transition metals in general. To our best knowledge, no hydride is experimentally known to date. Recent developments in the field of ultra-high pressures, in particular the development of the diamond-anvil cell (DAC) techniques, allows for synthesis and detailed characterization of new species, including hydrides, at unprecedentedly high pressures [17]. Nevertheless, while the DAC techniques become increasingly available they are still far from routine and, particularly in the case hydrogen-rich compounds, require significant amount of expertise and experimental effort.

In the absence of experimental results we thus apply modern computational techniques to explore the phase diagram of the $Ir/H_2$ system in a wide pressure range, from 1 atm. to 125 GPa, and search for hitherto unknown iridium hydrides. We analyze the possibility of high-pressure formation and stabilization of several stoichiometric iridium hydrides including $Ir_2H$, IrH, $IrH_2$, $IrH_3$, $IrH_4$ and $IrH_6$. For each stoichiometry we performed an extensive search for the most stable structures at 25 (low-pressure limit) and 125 GPa (high-pressure limit) combining several complementary approaches, including application of modern evolutionary algorithms [18-19] which are based on purely chemical composition (with no experimental input required).

At the level of our calculations we predict that iridium should react with excess hydrogen already at relatively low pressures of P > 5 GPa. Interestingly, at the lowest considered pressures, our calculations predict enhanced stability of a trihydride, $IrH_3$. Only at pressures higher than 75 GPa, a dihydride, $IrH_2$, stabilizes to the degree that it eventually becomes the preferred stoichiometry/phase. At the same time, our calculations predict that a monohydride should be thermodynamically highly unstable with respect to decomposition into pure elements. We also note that the proposed hydrogen-rich $IrH_3$ phase is the first stable trihydride predicted for platinum-metal group hydride.

## II. METHOD

All of the reported calculations were performed using density functional theory (DFT) with the Perdew–Burke–Ernzerhof (PBE) [20-21] parameterizations of the generalized gradient approximation (GGA) as implemented in the VASP code (ver. 4.6) [22-23]. The projector-augmented wave (PAW) method [24-25] was applied with PAW pseudo-potentials taken from the VASP archive. For the plane-wave basis-set expansion, an energy cutoff of 650 eV was used. Valence electrons (Ir: $6s^2 5d^7$; H: $1s^1$) have been treated explicitly, while standard VASP pseudopotentials (also accounting for scalar relativistic effects) were used for the description of core electrons. The Γ-centered *k*-point mesh was generated for every structure in such a way that the spacing between the *k* points was approximately 0.10 Å. Structures have been optimized using a conjugate-gradient algorithm with a convergence criterion of $10^{-7}$ eV. This ensured that in the optimized structures forces acting on the ions were in the order of 1 meV Å$^{-1}$ or smaller. The enthalpies derived from our calculations do not contain the zero-point-energy (ZPE) corrections.

For each stoichiometry we performed an extensive computational search for the most stable structures combining several complementary approaches: prototypical structures, purely random structure generation and evolutionary algorithms. In our calculations we used up to Z=4 formula units per unit cell. Conveniently, for most of the structures, except $IrH_2$, the predicted low- and the high-pressure ground-state structures are effectively identical. In the following analysis we focus on the thermodynamically best candidate structures only. For the particular case of $IrH_2$ we consider two competitive phases. We note that at pressures higher than 125 GPa more stable phases can, and do emerge; we chose not to investigate those further in this work.

Dynamical stability of the enthalpically preferred iridium hydride structures has been assessed through phonon analysis, within the harmonic approximation. Phonon dispersion curves were calculated using the finite-displacement method as implemented in the PHON code [26]. We used sufficiently large supercells, typically in the order of 3×3×3, in the interpolation of the force constants required for the accurate phonon calculations.

Finally, we stress in particular that the structures we consider are all ground state arrangements whereas the measurements are carried out at finite, typically close to room, temperatures.

## III. RESULTS AND DISCUSSION

### A. Enthalpy of formation; the tie-line plot

We searched for iridium hydride ground state structures for a large number of predefined Ir:H ratios (2:1, 1:1, 1:2, 1:3, 1:4, 1:6) and in a wide, 1 atm. < P < 125 GPa, pressure range.

In Fig. 1 we present the enthalpy of formation of $IrH_2$ and $IrH_3$, calculated with respect to pure elements, given as function of external pressure.

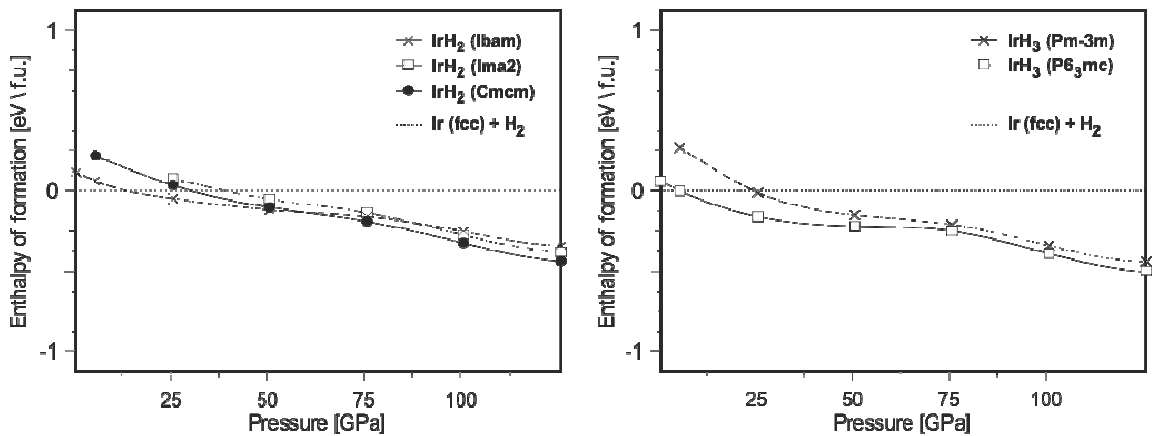

**Fig. 1**. Calculated enthalpy of formation for $IrH_2$ and $IrH_3$. The thermodynamic stabilization pressure of $IrH_2$ (Ibam) and $IrH_3$ ($P6_3mc$) is calculated to ~5 GPa and ~14 GPa, respectively.

For $IrH_2$, a low- and high-pressure phase as well as one additional potentially competitive phase are included in the plot; in this particular case, a phase transition from *Ibam* to *Cmcm* symmetry is expected to occur at approx. 60 GPa. For $IrH_3$ the low-pressure phase remains identical with the predicted high-pressure one and is calculated to be stable until at least 125 GPa. A third most stable phase is also included in the plot for comparison.

To help determine which of the stoichiometries actually belong to the $Ir/H_2$ phase diagram at a given pressure, the enthalpies of formation of the identified ground state structures were calculated at several pressures with respect to pure iridium and pure hydrogen in their most stable forms (*fcc* for iridium, in the whole pressure range, and depending on the applied pressure, P63/m or C2/c for hydrogen) [27].

We analyzed the relative thermodynamic stability of each stoichiometry by plotting the enthalpies of formation, here calculated per atom, as a function of hydrogen content. The resulting tie-line plot is given in Fig. 2. Those of hydrides which are stable with respect to disproportionation into *other* hydrides and/or pure elements are those forming a convex hull of energy with respect to composition.

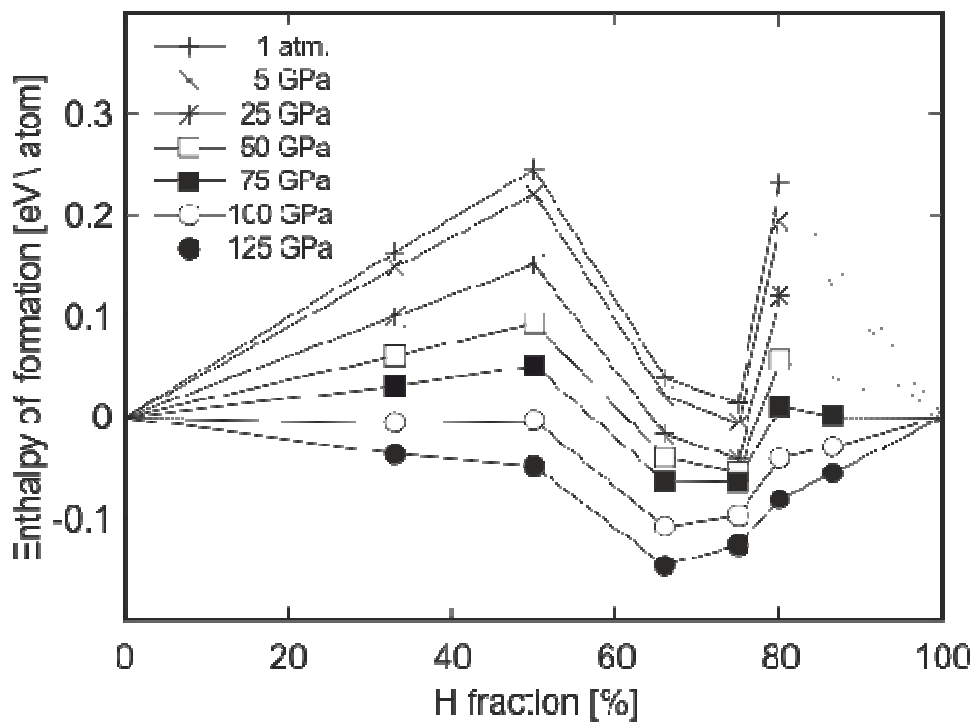

**Fig. 2**. The relative enthalpies of formation (per atom) of the $IrH_x$ phases calculated with respect to elemental iridium and molecular hydrogen. Hydrides which are stable with respect to disproportionation are those forming a convex hull of enthalpy with respect to composition into selected species.

As the pressure increases, the enthalpy of formation of each hydride becomes less and less positive and/or more and more negative, proving that a pressure increase is likely to stabilize iridium hydrides.

At P=1 atm. all the compositions studied appear unstable with respect to disproportionation into pure elements. This confirms the experimentally observed lack of hydride formation at ambient conditions. With pressure increase, all of the considered hydrides stabilize but two of the stoichiometries, $IrH_2$ and $IrH_3$, clearly stand out.

At P~5 GPa, $IrH_3$ is the first hydride to stabilize thermodynamically with respect to disproportionation into pure elements, although significant imaginary frequencies are found in a phonon analysis. Above 25 GPa the imaginary frequencies are no longer present, the structure ($P6_3mc$) becomes dynamically stable and remains so until at least 125 GPa.

At ~14 GPa also a dihydride stabilizes thermodynamically with respect to disproportionation into the elements but its relative stability remains lower than that at the trihydride (see the tie-line plot, Fig. 2). Calculations indicate that it requires pressure as high as 75 GPa for the dihydride to stabilize over the trihydride. This behavior is somewhat counterintuitive as one would rather expect the opposite trend; initial stabilization of a dihydride (or perhaps even a monohydride) followed by subsequent stabilization of higher hydrides. However, further structural analysis reveals that $IrH_2$ and $IrH_3$ should not be considered as simple interstitial hydrides in which the metal host lattice progressively saturates with interstitial H atoms.

Finally, we note that these purely enthalpic arguments do not include kinetic factors, such as the rate of diffusion of hydrogen into the metal lattice, nor the possibility of high potential energy barriers for reactions of Ir with $H_2$. In addition, we repeat that the arguments formally apply only to ground state conditions. This leads to a further necessary caution that the zero-point energies of the structures have not been included in the tie-line plot of Fig. 2 from which our conclusions are derived. Inclusion of those might favor the other competitive phases and/or shift the stabilization pressure to higher absolute values. Nonetheless, we still predict that at sufficiently high pressure (in the order of 20-25 GPa) and in the environment of excess hydrogen, $IrH_x$ *should* form, with *x* substantially higher than 1.

We now proceed with the detailed structural characterization of the predicted most stable phases.

**B. Structures**

Bulk iridium adopts the cubic face-centered structure and maintains it in the whole considered pressure range. Our reference calculations found that the next most competitive structure corresponds to *hcp* structure separated from the *fcc*, on average, by -0.07 eV/atom. The difference increase as the pressure grows higher.

Concerning the hydrides, there are certain trends one might expect. Obviously an increase of pressure should bring all Ir and H atoms closer together, at least on average. Depending on the type of interactions the Ir-H "bonds" (or more accurately, Ir-H contacts) should be more or less susceptible to compression. Aside from that, since we are mainly on the hydrogen-rich side of the phase diagram, the number of hydrogens around a statistical iridium atom should increase as one goes from $Ir_2H$ to $IrH_6$ at a given pressure although at certain point one should start to observe "loose" $H_2$ *molecules*, particularly at low pressures and for high Ir:H ratios. With those expectation in mind let us now look at the ground state structure of each of the hydrides.

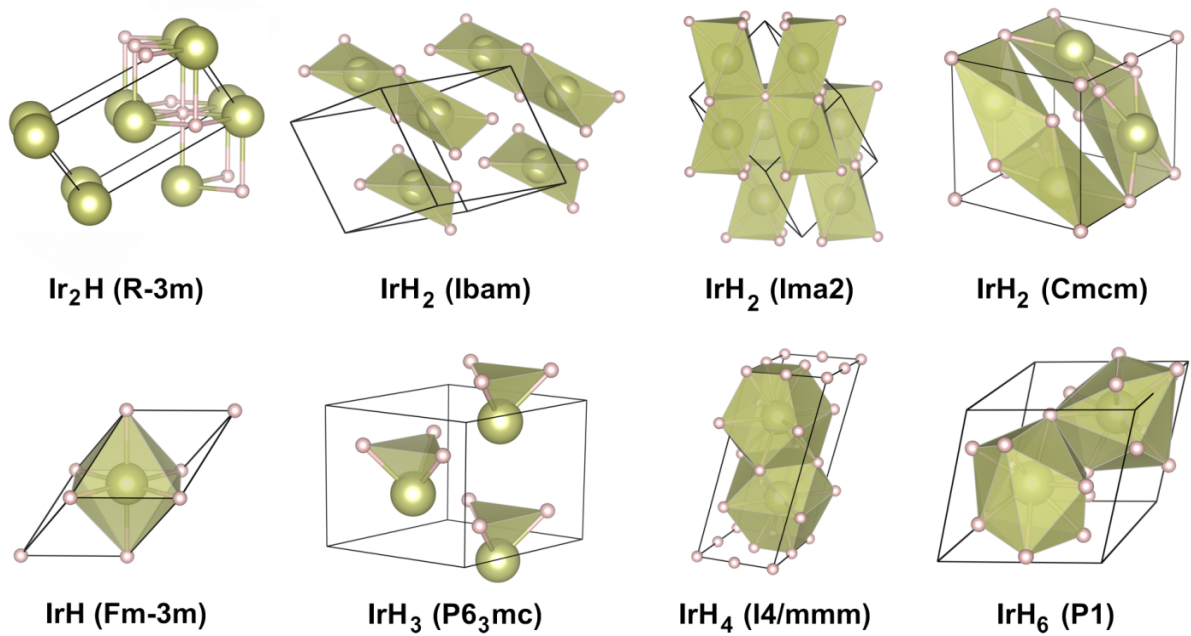

**Fig. 3**. Schematic presentation of the unit cells of the most stable phases (Ir: gold atoms, H: pink atoms).

## 1. Ir$_2$H and IrH

The two hydrides, Ir$_2$H and IrH, share a common feature: in both cases the metal frameworks form, effectively, an *fcc* sublattices in which hydrogen atoms occupy either one or two octahedral sites.

Note that for the *fcc* arrangement, one octahedral site is available *per* metal atom. Thus, in the case of Ir$_2$H only one such sites is occupied, resulting in lowering of the metal sublattice symmetry to *R-3m* (rhombohedral). The distortion to the metal sublattice caused by the presence of the octahedral vacancy remains relatively small and as such we considered it as only a minor departure from the *Fm-3m* symmetry of bulk iridium. See Fig. 3 for the actual unit cell.

The lowest enthalpy structure found for the equimolar stoichiometry of iridium and hydrogen is the rock-salt structure, of *Fm-3m* symmetry. In this structure iridium atoms form a perfect *fcc* network in which hydrogen atoms occupy all the available octahedral interstitial sites. The unit cell for this system is shown in Fig. 3.

At 25 GPa the calculated shortest Ir-Ir contact in Ir$_2$H is 2.752 Å and the shortest Ir-H contact is 1.978 Å (here we refer to the close-neighbor environment of the octahedrally coordinated H atom). The analogous distances for the vicinity of the octahedral vacancy (oct) are: Ir-Ir = 2.682 Å and Ir-oct = 1.921 Å (as expected, both the distances are shorter due to the missing interstitial

H atom). In the case of IrH the analogous shortest Ir-Ir and Ir-H bonds are 2.878 and 2.035 Å, respectively. At the same pressure, those distances are only slightly longer than in pure iridium (2.679 and 1.894 Å, respectively) meaning that the two hypothetical hydrides should be considered as simple interstitial species. We emphasize that the shortest Ir-H distance is in the order of 2.0 Å, suggesting lack of significant covalent Ir/H interaction.

Finally, we reiterate that from purely thermodynamical point of view both the stoichiometries are *highly* disfavored (see Fig. 2). This is in accord with the findings of Antonov *et. al.* who observed an extremely low hydrogen uptake by bulk iridium (x=0.005 in IrH$_x$). At any given pressure, Ir$_2$H and IrH are found to be less stable than other, higher hydrides.

**2. IrH$_2$**

Our calculations yield several competitive phases for the 1:2 ratio. In this report, we focus on two selected phases, a low-pressure ground state *(Ibam)* and a high-pressure modification *(Cmcm)*. The two structures are depicted in Fig. 3. In Table 1 we summarize selected structural parameters, including the number of close-contacts and the corresponding bond lengths calculated at two pressure points.

**Table 1.** Selected bond lengths (A), number of Ir-H contacts, calculated for selected phases of IrH$_2$ (Ibam) at 75 and 125 GPa.

| P | Ir-Ir (eff. *sh*) | H---H | Ir-H | No. of contacts (Ir-H < 1.9 A) |
|---|---|---|---|---|
| Ibam @ 75 GPa | 2.650 (x6) <br> 2.731 (x2) | 1.642 | 1.674 (x4) <br> 2.214 (x4) | 4 |
| Cmcm @ 75 GPa | 2.642 ÷ 2.682 (x6) <br> 2.726 (x2) | 1.542 | 1.701 (x2) <br> 1.747 (x1) <br> 1.855 (x2) <br> 2.040 (x1) <br> 2.114 (x2) | 3+2 |
| Ibam @ 125 GPa | 2.578 (x6) <br> 2.651 (x2) | 1.556 | 1.659 (x4) <br> 2.115 (x4) | 4 |
| Cmcm @ 125 GPa | 2.576 ÷ 2.608 (x6) <br> 2.659 (x2) | 1.481 | 1.676 (x2) <br> 1.720 (x1) <br> 1.812 (x2) <br> 1.967 (x1) <br> 2.036 (x2) | 3+2 |

[a] selected shortest Ir-H distances, [b] shortest H---H separation

In both the cases, iridium sublattice formally adopts orthorhombic symmetry. Close inspection reveals that the effective arrangement of the metal atoms actually corresponds to simple-hexagonal packing scheme (see Fig. 3 and Table 1). In fact, at any given pressure the metal frameworks of *Ibam* and *Cmcm* phase are effectively identical (compare the Ir-Ir distances in Table 1). The minor differences arise from different hydrogen packing. This alternative hydrogen packing also results in different coordination environment of the central iridium atoms eventually yielding the two enthalpically distinct phases: the *Ibam* and *Cmcm*.

In the *Ibam* phase, iridium is tetra-coordinated with all the Ir-H contacts full symmetrised. At 75 GPa the shortest Ir-H bond length is 1.674 Å, while the second-close Ir-H distance is 2.214 Å. Interestingly, at 125 GPa the shortest Ir-H bond is 1.659 Å (nearly unaffected by extra pressure) while the second-close Ir-H contact shortens substantially to 2.112 Å. Detailed structural analysis reveals that the *Ibam* phase consists of 1-dimensional (1D) chains in which each hydrogen acts as a bridging atom shared by two iridium atoms. We conclude that within each such chain hydrogen is primarily covalently bonded with iridium resulting in stiff bonds. Introduction of external pressure thus affects primarily the inter-chain distance, as shown in Table 1.

In the *Cmcm* phase, each iridium atom is now five-coordinated although the Ir-H distances vary from 1.701 to 1.855 Å (at 75 GPa). The actual coordination environment is thus more accurately described as 3+2, with three short and two substantially longer bonds (see Table 1). The length of the three shortest Ir-H bonds remains essentially indifferent with pressure increase; each H atom bridges two iridium atoms. In case of the remaining two longer Ir-H contacts the hydrogen atoms are shared equally between *four* metal atoms. Those contacts are substantially more susceptible to pressure-dependent variation in length. We conclude that in the *Cmcm* structure, the three bridging hydrogen atoms should be considered as primarily covalently bonded while the remaining two (Ir-H > 1.8 Å) as interstitial atoms.

### 3. IrH$_3$

Selected results on the number of close-contacts and the corresponding bond lengths calculated for the ground-state IrH$_3$ structure are summarized in Table 2.

According to our simulations the most stable phase of IrH$_3$ adopts the *P6$_3$mc* structure shown in Fig. 3. In it, metal atoms occupy positions corresponding to *hcp* lattice; each iridium is now six-coordinated by neighboring metal atoms. The shortest Ir-Ir distance, calculated at 75 GPa, is 2.691 A and is effectively identical as the shortest Ir-Ir distance calculated for the two previously discussed IrH$_2$ phases (2.691 Å vs 2.650/2.578 Å for *P6$_3$mc* vs *Ibam/Cmcm*, respectively).

Interestingly, each iridium atom is now coordinated by *three* hydrogen atoms. The resulting three Ir-H bonds are short, in the order of 1.7 Å. Also the H-Ir-H angle varies little with pressure, from 77.0° to 73.8° and 77.7° at 25, 75 and 125 GPa, respectively. At 25 GPa each hydrogen is exclusively connected with one iridium atom only. Those bonds should thus be considered primarily covalent. Note also that the second-closest Ir-H contacts are substantially longer, particularly at low-pressure limit, and are prone to shortening with pressure increase (see Table 2). In fact, the short Ir-H bonds actually lengthens slightly with pressure; the three bonds weaken while six new are being formed. We suggest that at the low-pressure limit the proposed

$P6_3mc$ phase of IrH$_3$ actually exhibit molecular character. Only with a substantial pressure increase all nine Ir-H contacts eventually nearly equalize forming an extended solid with each hydrogen now effectively shared between two iridium atoms (125 GPa).

**Table 2.** Selected bond lengths (A), number of Ir-H contacts, calculated for IrH$_3$ (P6$_3$mc) at 25, 75 and 125 GPa.

| P | Ir-Ir[a] (eff. *hcp*) | H---H[b] | Ir-H[a] | No. of contacts (Ir-H < 1.9 A) |
|---|---|---|---|---|
| P63mc @ 25 GPa | 2.840 (x6) 3.648 (x2) | 1.693 | 1.666 (x3) 2.001 (x6) | 3 |
| P63mc @ 75 GPa | 2.705 (x6) 3.496 (x2) | 1.521 | 1.711 (x3) 1.866 (x6) | 3+6 |
| P63mc @ 125 GPa | 2.621 (x6) 3.361 (x2) | 1.419 | 1.722 (x3) 1.806 (x6) | 3+6 |

[a] selected shortest Ir-Ir, Ir-H distances, [b] shortest H---H separation.

### 4. IrH$_4$ and IrH$_6$

Both the IrH$_4$ and IrH$_6$ are thermodynamically unstable with respect to pure elements until approx. 75 GPa. Even at the highest considered pressures both the ratios remain highly unstable with respect to disproportionation into hydrogen and lower hydrides. Thus, none of those hydrides is expected to form, not even under the most extreme of the considered conditions. We thus characterize the most stable identified phases only briefly.

The preferred structure found for IrH$_4$ is the *I4/mmm* structure shown in Fig. 3. The metal sublattice adopts a relatively complex tetragonal symmetry. Each Ir atom is now coordinated by 11 H atoms. In contrast to IrH$_2$ and IrH$_3$ the shortest and the longest Ir-H distance is now 1.84 Å and 1.94 Å, respectively (at 125 GPa). This structure exhibit features typical for an interstitial hydride. We speculate that the apparent lack of covalent interactions leads to its large thermodynamic instability.

The ground state structure found for IrH$_6$ exhibit no symmetry at all. The metal sublattice alone does have C2 symmetry but since the metal framework is filled with hydrogen atoms in mostly random manner no symmetry is eventually observed.

Finally, it has to be noted that in the cases of IrH$_6$ there is a limiting pressure of approx. 75 GPa below which the identified thermodynamically most stable structures can only be described as mixtures of molecular H$_2$ and hydrides of lower stoichiometry.

## C. Dynamic stability

Phonon calculations are of interest both as guarantors of dynamical stability and with respect to the information such calculations may yield on the ease or difficulty of motions in the structures considered. We have performed phonon analysis at two pressure points: at 25 and 125 GPa, corresponding to low and high pressure limits. The analysis reveals that all the reported phases are dynamically stable within the studied pressure range. At P < 25 GPa we did observe negative frequencies for $IrH_3$ but those quickly stabilize upon pressure increase. Also, the higher hydrides, $IrH_4$ and $IrH_6$ are dynamically stable only at the upper limit of the considered pressure range.

For $Ir_2H$ and $IrH$, two distinct phonon branches can be observed (see Supplementary Material, Fig. S1 and S2). At 125 GPa the gap between the two groups of vibrations is rather wide, of the order of 600 $cm^{-1}$. It is as if the hydrogens and iridium atoms moved separately. Indeed, this is what we saw upon visualization of the particular modes. It also agrees with the conclusion resulting from the structural analysis on the interstitial nature of the two hydrides.

We present three plots, calculated for $IrH_2$ and $IrH_3$ illustrating phonon dispersions and the corresponding phonon DOS (PhDOS) at 125 GPa; see Fig. 4, 5 and 6.

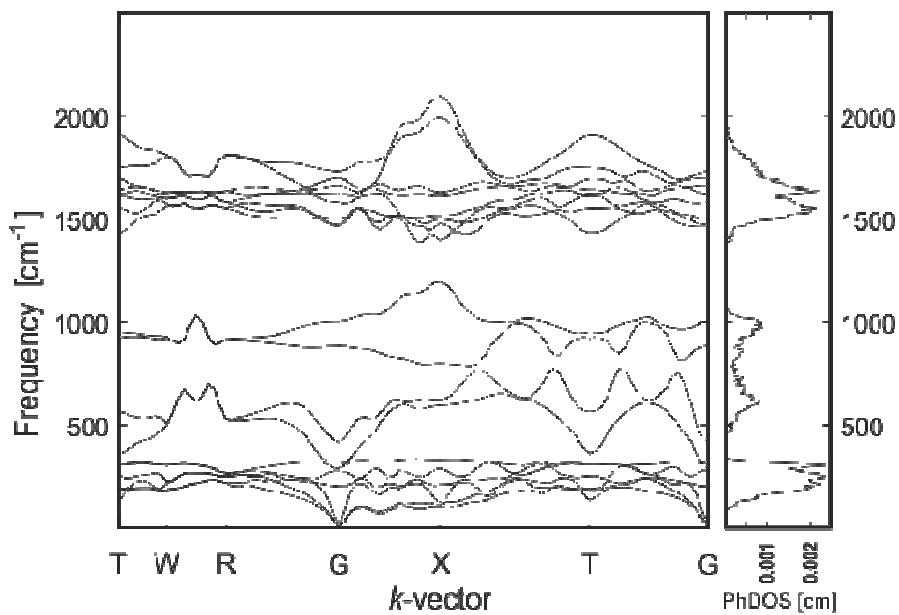

Fig. 4. Phonon dispersions and phonon density of states, PhDOS, calculated for **$IrH_2$ (Ibam)** at 125 GPa.

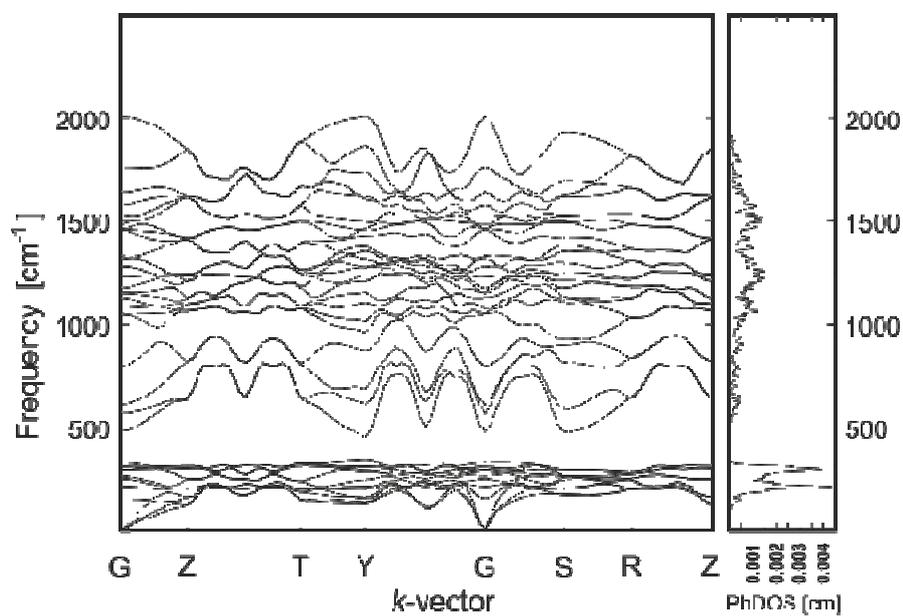

Fig. 5. Phonon dispersions and phonon density of states, PhDOS, calculated for **IrH$_2$ (Cmcm)** at 125 GPa.

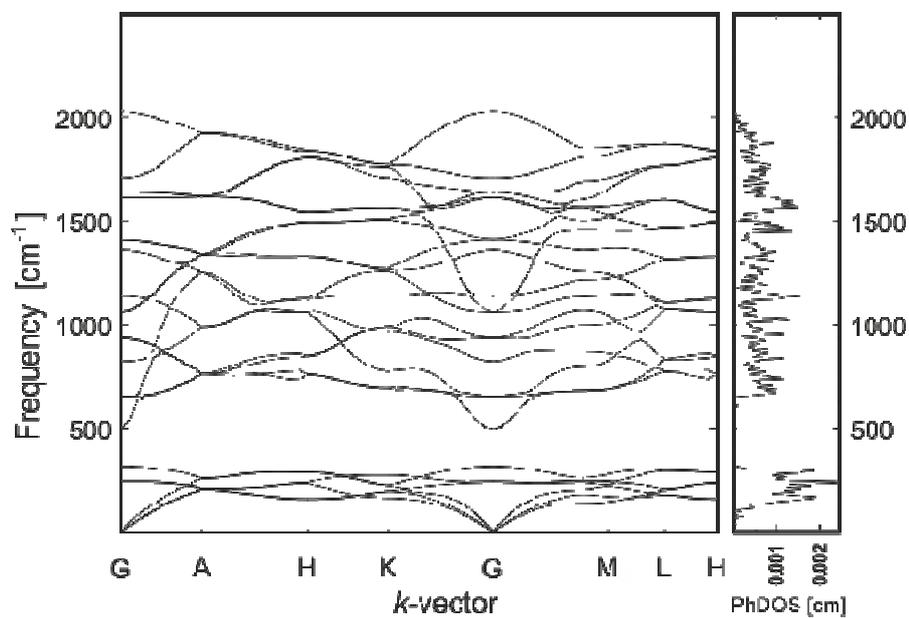

Fig. 6. Phonon dispersions and phonon density of states, PhDOS, calculated for **IrH$_3$ (P6$_3$mc)** at 125 GPa.

For IrH$_2$ *(Ibam)* the phonon branches separate into three groups. The high-energy phonons are attributed to Ir-H stretching modes, the medium-energy phonons are attributed to Ir-H bending modes and the low-energy modes arise from metal sublattice vibrations only.

In case of IrH$_2$ *(Cmcm)*, the low-energy modes (up to 300 cm$^{-1}$) arise, once again, mainly from the movement of heavy iridium atoms. Bending and stretching modes of Ir-H contribute to phonon spectrum in the range of 500-1000 and 1500-2000 cm$^{-1}$, respectively. The intermediate phonons in the range of 1000-1500 arise mainly due to the presence of the interstitial hydrogen atoms moving much more independently with respect to the metal framework.

For IrH$_3$ (P6$_3$mc) one can observe a clear separation of low-energy modes arising due to metal framework librations. The remaining modes, in the 500-2000 cm$^{-1}$ range, form a continuous spectrum with no differentiation into separate phonon branches. The lack of clear differentiation is attributed to the fact that at 125 GPa all the Ir-H bonds already equalize and the hydrogen atoms are no longer exclusively bound to one iridium but rather act as bridging atoms between two instead. A closer inspection reveal that all those modes correspond to numerous Ir-H-Ir stretching and bending modes.

It is interesting to compare the phonon spectra of IrH$_x$, $x$=1–6, with those we computed recently for PbH$_4$ under pressure [28]. The Ir and Pb systems share some features: a mass disparity between the heavy atoms and the hydrogens and a common instability at P=1 atm with respect to the elements. One marked difference is that for IrH$_x$ we *did not* find H$_2$ units at all, whereas for PbH$_4$ we do find them, albeit stretched.

**D. Electronic properties**

To analyze the conductive properties of the iridium hydrides we calculate band structures and corresponding electronic densities of states for the optimized ground state structures. Here, we note that the Kohn–Sham formulation of DFT is well known to underestimate the width of electronic bandgaps. While we did not try to correct for this the study of the electronic structures is still likely to give us reliable, qualitative information about the insulating-semiconducting-metallic character of the considered iridium hydrides phases. We here focus on characterization of the two viable stoichiometries, IrH$_2$ and IrH$_3$.

**1. IrH$_2$**

The dihydride is expected to be the dominant stoichiometry only at P > 75 GPa at which point the *Cmcm* phase is expected to form. Nevertheless, here we report on the band structure and the DOS of the low-pressure *(Ibam)* and the high-pressure phase *(Cmcm),* both calculated at 125 GPa for consistency (see SM: Fig. S6 and Fig. S7 for results calculated at lower pressures).

The results are shown in Fig. 7 and 8. Analysis of the DOS indicates accumulation of occupied states at the Fermi level ($E_f$) and is indicative of metallic character. Even at the lower pressures, the density of states at $E_f$ remains substantial. Additional analysis of the DOS partitioned with respect to individual atomic contributions reveals significant mixing of the hydrogen *s* and metal *d* states, particularly at lower energies. The closer the Fermi level the less mixing is observed and at the Fermi level the major contribution is due to the occupied 5d states of iridium atoms only. Note that for the two phases, *Ibam* and *Cmcm*, the calculated DOS remains nearly identical. This is attributed to the fact, that any given pressure the metal frameworks of the two phases are effectively identical.

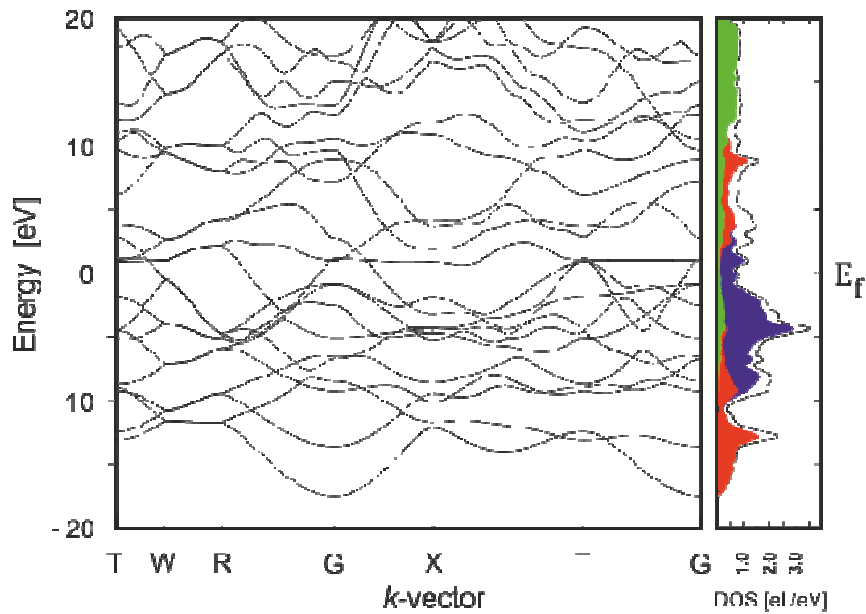

**Fig. 7.** Total electronic density of states (DOS) per valence electron calculated for $IrH_2$ (Ibam) at 125 GPa (red: s-character, green: p-character, blue: d-character, black: total DOS). The Fermi energy ($E_f$) is set to zero for convenience.

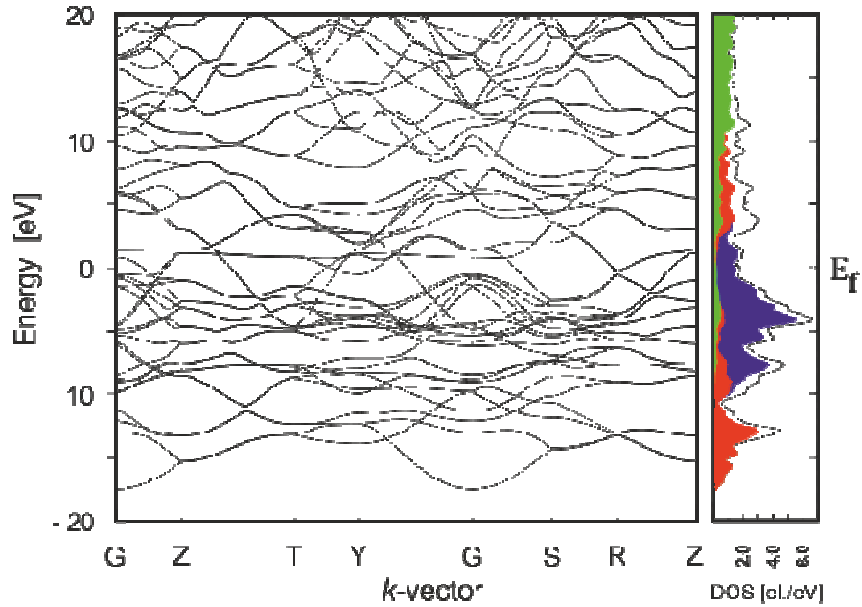

**Fig. 8.** Total electronic density of states (DOS) per valence electron calculated for **IrH$_2$ (Cmcm)** at 125 GPa (red: s-character, green: p-character, blue: d-character, black: total DOS). The Fermi energy ($E_f$) is set to zero for convenience.

### 2. IrH$_3$

The IrH$_3$ is expected to be the dominant stoichiometry at 25 GPa < P < 75 GPa. Here we present the results calculated for P=25 GPa only (see SM: Fig S8 for results calculated at other pressures).

In case of the trihydride (P6$_3$mc) we observe a deep pseudogap in the calculated DOS located at the Fermi level. A close investigation of the band structure reveals that at P=25 GPa none of the bands actually cross the Fermi level (see Fig. 9). Thus, at low pressure limit the trihydride is expected to be a semiconductor (the actual calculated width of the band gap at 25 GPa is 0.13 eV). At higher pressures the band gap eventually closes and the bands start to cross the $E_f$ but since the accumulation of the occupied states remains low the trihydride should be considered only a weak metal (a semimetal) at best .

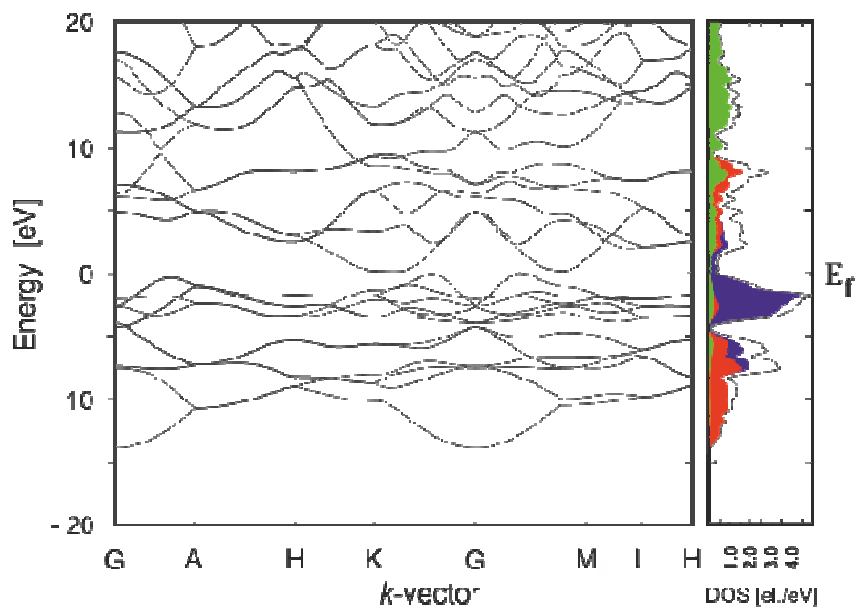

**Fig. 9.** The partial electronic density of states (PDOS) per valence electron calculated for **IrH$_3$ (P6$_3$mc)** at 25 GPa (red: s-character, green: p-character, blue: d-character, black: total DOS). The Fermi energy (E$_f$) is set to zero for convenience.

As in the case of dihydrides one observes significant mixing of the hydrogen *s* states with the metal *d* states at lower energies but, in contrast to the dihydirdes, the higher energy *d* states are now well separated from the s/d mixture by a pseudogap located at 12 el. count.

The lack of pronounced metallic character indirectly supports our claim that the trihydride might exhibit molecular character, particularly at 25 < P < 75 GPa.

**E. Final word on bonding**

In both IrH$_2$ and IrH$_3$, iridium is expected to form primarily covalent bonds with hydrogen. To complete the description of the predicted hydrides and further confirm the covalent nature of the Ir-H bonds in IrH$_2$ and in particular, IrH$_3$ we plot the electron density. For consistency we chose the same pressure and the same cut-off value for all the cases. The results are shown in Fig. 10. Note the presence of increased electron density between iridium and hydrogen atoms. Also, in the case of IrH$_2$ *(Cmcm)* note the presence of hydrogen with spherical electron density distribution, indicative of interstitial character of the guest atom.

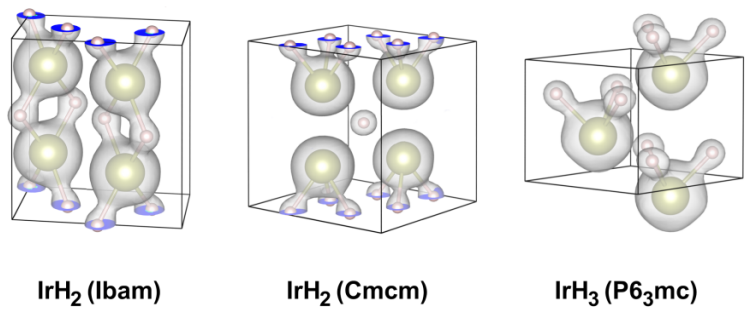

Fig. 10. The structure of IrH$_2$ (Ibam), IrH$_2$ (Cmcm) and IrH$_3$ (P62m) with superimposed electron density isosurface (contour cut-off @ 0.12).

In the predicted di- and trihydrides the shortest distances between two hydrogen atoms, H---H, are short and comparable with the second-shortest distance in pure, molecular H$_2$ [29]. The H---H separation calculated for H$_2$ is 1.59 and 1.33 Å at 75 and 125 GPa, respectively (see Table 1 and 2 for comparison). Switendick suggested that for a stable interstitial hydride to form, the shortest H---H separation must be larger than 2.1 Å [30]. In all three cases the predicted H---H distances are substantially shorter, suggesting that the proposed hydrides form a new class, of primarily covalently bound hydrides, among platinum-metal group and transition metal in general.

## IV. CONCLUSIONS

All the considered phases were calculated to be thermodynamically unstable with respect to separation into the elements at P=1 atm. At higher pressures a variety of stoichiometries stabilize. Two stoichiometries, IrH$_2$ and IrH$_3$, are found to be particularly stable and as such are expected to form spontaneously at elevated pressure and in the excess hydrogen environment. An interstitial monohydride *(Fm-3m)* is found to be particularly unstable.

For the first time, a stable trihydride is predicted within the platinum-metal group. The trihydride (IrH$_3$: *P6$_3$mc*) is computed to stabilize thermodynamically at ~5 GPa although its dynamical stabilization is expected at P>25 GPa. The thermodynamic stabilization of a dihydride (IrH$_2$: *Ibam*) is calculated at ~14 GPa and coincide with its dynamical stabilization. Within the applied level of theory, the 1:3 ratio is expected to be thermodynamically preferred one until at least 75 GPa at which point 1:2 ratio takes over.

In both IrH$_2$ and IrH$_3$, iridium is expected to form covalent bonds with hydrogen. Application of Pyykko's single-bond covalent radii yields an Ir-H bond distance in the order of 1.55 Å. The average Ir-H distance of the expected covalent bonds in IrH$_2$ and IrH$_3$ is in the order of 1.65-1.70 Å (at 75 GPa). We also note that those bond are resistant to compression.

The dihydride is expected to be a metal (with the metallic character mainly due to the metal sublattice). On contrary, due to molecular character of IrH$_3$, the trihydride is expected to be a semiconductor or, particularly at higher pressures, a semimetal.

Finally, we note that in all three cases the metal framework of the predicted tri- and dihydrides is a clear departure from the *fcc* structure of bulk iridium; in case of the dihydride the metal sublattice adopts an orthorhombic symmetry (effectively simple-hexagonal) and in the case of trihydride, a hexagonal-close-packed. Should one try to synthesize the hydrides, such dramatic symmetry changes should help identify the resulting species.

**Acknowledgements**

The author thank prof. Roald Hoffmann, prof. Pekka Pyykkö and prof. Marek Tkacz for comments and discussion. The grant UMO-2012/05/B/ST3/02467 of the Polish National Science Centre (NCN) is acknowledged.